# A SURVEY OF GALAXY REDSHIFTS FROM LOW-RESOLUTION SLITLESS SPECTRA: SUGGESTION OF LARGE-SCALE STRUCTURES AT Z ~0.1

M. Beauchemin[1,2,3] and E. F. Borra[1,2]

Centre d'Optique, Photonique et Laser
Département de Physique, Université Laval,
Québec, QC
CANADA G1K 7P4

POSTAL ADDRESS: Département de Physique, Université Laval, Québec, QC, Canada G1K 7P4

[1] Guest Observers, Canada France Hawaii Telescope
[2] Also Observatoire du Mont Mégantic
[3] Now at the Département des sciences géodésiques et télédétection, Université Laval, Québec, QC, Canada G1K 7P4

astro-ph/9408031   10 Aug 94

## ABSTRACT

We have conducted a galaxy survey based on low-resolution slitless spectra taken from the automated CFHT-Laval survey. We present redshift distributions for 522 galaxies distributed in 4 distinct regions of the sky. Redshifts are determined from the shifted positions of the 4000 Å stellar break using an automated break-finding algorithm. The redshifts so obtained have a precision of $< 3000$ km s$^{-1}$, good enough to trace the large-scale distribution of galaxies. The most striking feature of the survey is an apparent excess of galaxies at $z \sim 0.1$ in 8 of the 10 fields probed. Although the statistical significance of the peaks appears marginal ($\sim 2.5\sigma$ per field) after taken in account the small-scale galaxy clustering, the peaks may nevertheless reveal the existence of large-scale structures at $z \sim 0.1$, indicating that the known large structures in the local universe are not unusual.

# 1 INTRODUCTION

A slitless spectroscopic survey based on photographic IIIa-J plates taken with the Canada-France-Hawaii telescope and the blue grens (grating-prism-lens combination) is currently being analyzed at Laval University (see Borra et al. 1987 for a general description). The main goal of the survey is the study of quasars (see Beauchemin & Borra 1991 and references therein). However, tools have also been developed to extract information from stars (Beauchemin, Borra & Levesque 1991) as well as galaxies (Borra & Brousseau 1988; hereafter referred to as BB).

Most existing redshift surveys of galaxies probe apparently bright ($B < 17$) or faint ($B > 20$) galaxies with little coverage in the $17 < B < 20$ magnitudes range. Moreover, bright surveys usually cover wide angles ($> 10$ deg$^2$) and faint ones are restricted to small surfaces ($< 0.3$ deg$^2$) (see Giovanelli & Haynes 1991 for a review). The useful range of magnitudes for galaxies studied on grens plates is about $16 < B < 20.5$ and a typical field of 3 plates spans 3 degrees and therefore fills a niche among existing surveys. BB have shown that it is possible to obtain redshifts of galaxies from low-resolution spectra to $z \sim 0.3$. The technique used to derive galaxy redshifts from grens spectra is fully described in BB. In low-resolution spectra, the redshift information is mainly available from the 4000 Å stellar break: it is prominent in the spectra of elliptical galaxies and Sa-Sb spirals but is weaker in later Hubble types (Coleman, Wu & Weedman 1980). The technique is based on a break-finding algorithm which locates the redshifted 4000 Å break. The galaxy redshifts so obtained have a precision of $\sim 3000$ km s$^{-1}$, good enough to trace the large-scale distribution of galaxies. Schuecker & Ott (1991) have also developed an automated method for the determination of galaxy redshifts for prism spectra with a quoted precision similar to ours. Note, however, that their spectral information is more 'diluted' than on our grens material. The prism plates have a reciprocal dispersion of 2480 Å mm$^{-1}$ at H$\gamma$, a plate scale of 67" mm$^{-1}$ and seeing disks ~ 3" while the grens plates have a dispersion of 1000 Å mm$^{-1}$, a scale of 14" mm$^{-1}$ and seeing disks of 1". Our redshifts should therefore have better precision.

The survey as it is now, includes 10 plates deep enough to give a sufficient number of galaxies for which we can obtain redshifts from break detections (~ 50 galaxies per rectangular areas of ~ 0.6 deg$^2$). In this paper, we present the redshift distributions for 522 galaxies in 4 distinct regions of the sky, for a total surface of ~ 5 deg$^2$. The paper is organized as follows: we describe in Section 2 the construction of the galaxy catalogue. We summarize in Section 3 the way redshifts are determined. The redshift distributions of galaxies are presented in Section 4 and the results are discussed in Sections 5 and 6. We will assume a Hubble constant of $H_o$=100 km s$^{-1}$ Mpc$^{-1}$.

## 2 THE GALAXY CATALOGUE

The automated CFHT-Laval survey is based on photographic IIIa-J plates obtained with the Canada-France-Hawaii telescope and the blue grens (Borra et al. 1987). The wavelength coverage is about 3400-5400 Å and the resolution, which depends on seeing as well as image extendedness, is ~ 75 Å. The typical signal-to-noise (s/n) ratio for the plates studied in this paper is ~ 10 per 100 Å bin at $B$ =19. The plates are entirely digitized and analyzed using computer programs. The image analysis software used to analyze the plates is described in Borra et al. (1987). The noteworthy aspects of the survey are that every energy distribution and parameter are calculated with their associated errors and that $B$ and $V$ photoelectric sequences are available for each plate.

The star/galaxy separation technique is fully described in BB and Borra et al. (1987). Briefly, the program uses a total ($m_t$) and a nuclear magnitude ($m_n$) computed for every object (inside apertures of 1.7 and 4.3 arcsecond diameters respectively). The technique is based on the fact that stars should line up along a line of slope 1 when total ($m_t$) and nuclear magnitudes ($m_n$) are plotted against each other (justification for this can be found in Edwards et al. 1985). The scatter on the left side of the slope 1 line gives an estimate of the standard deviation due to photometric errors. The extendedness of an object is then given by its distance to the slope 1 line in units of standard deviations. We call this quantity extension class ($EC$). The compromise between star contamination and completeness in the selected galaxy catalogue depends on the value adopted for the extension class parameter ($EC$) which measures the distance, in standard deviation units, of an object from the locus of stars in a $m_t$ vs $m_n$ plot. To determine the $EC$ threshold for galaxy identification without including too many stars, we first select all objects with $B < 20$ and $EC > 3$ (spectral breaks on grens plates are detected to $B \sim 20$). Then, we inspect every object visually on the grens plates. Eyeball on-plate inspection is quite reliable because of the good plate scale (14 arcsec mm$^{-1}$) and because galaxy spectra with visible discontinuities have enough flux to see their extendedness. We then plot the eyeball classification in an $EC$ vs $m_n$ diagram obtaining a clear-cut separation among visually confirmed stars and galaxies, so that a threshold on $EC$ to identify galaxies is easily set. The threshold is typically $EC = 5$ and varies little from plate to plate. We can estimate the rate of star contamination in the resulting galaxy sample using the $EC$ distribution of *bona fide* stars on a plate. Spectra with a break detected at 4000 Å are obviously stars and will be considered as *bona fide* stars. The number of *bona fide* stars classed with $EC > 5$ compared to the total number of *bona fide* stars ($EC \geq 0$) gives the rate of star contamination in the galaxies locus ($EC > 5$). For all plates, we find that less than 5% of the total number of stars are actually misclassified as galaxies ($EC > 5$) by the software.

Additional visual checks on the grens plates are performed for each galaxy candidate and objects with stellar-like spatial profile are rejected from the list. About 10% of the galaxy candidates are borderline objects as their extendedness is not fully convincing after eyeball on-plate evaluation. The status of these borderline objects is checked independently using an independent catalogue constructed from COSMOS scans of UK Schmidt IIIa-J plates (Heydon-Dumbleton, Collins & MacGillivray 1989). By establishing the correspondence between Schmidt and grens positions, COSMOS galaxies are identified on grens plates. By intersecting the lists, we found that ~80% of all borderline objects, typically 4 out of 5 per plate, are classed as galaxies in the COSMOS catalogue. Visually inspected objects with borderline status are kept in the final catalogue only if they are also contained in the COSMOS catalogue. Because of technical problems, we do not have COSMOS classifications for H6; galaxies with borderline status on that particular plate have not been included in the final catalogue.

Completeness in the catalogue depends on break detection at a given magnitude. For the F, SGP2-3 and H plates, nearly 100% of all galaxies in the catalogue with $B < 19$ have a break detected, these numbers fall to 75% for $19 < B < 19.5$, 50% for $19.5 < B < 20.0$ and 25% for $20 < B < 20.5$. For the remaining plates, the statistics are the same but for magnitudes 0.5 mag brighter.

Table 1 lists the fields studied in this paper. These plates have been selected because they are the deepest ones. They are located in 4 distinct regions of the sky with 3 plates per region. The NGP region has only 1 plate. The area used on each plate is about 0.5 deg$^2$. There are typically ~ 50 galaxies per plate with detected breaks.

## 3  REDSHIFT MEASUREMENTS

The technique used to find breaks is described in BB. Briefly, the break-finding algorithm is based on the flux difference between two nearby spectral bands, each of 175Å width, which are moved across the spectrum in 25 Å steps. This gives a quantity, called $C_j$, which is proportional to the first derivative of the spectrum at position $j$. The nominal position of a spectral break is given by the point at which the second derivative of the spectrum is zero, that is, the point at which the slope changes sign. Since we know the uncertainties associated with every pixels, we can assert the statistical significance of $C_j$ via the associated s/n ratio ($\Sigma$). A uniform threshold of $\Sigma \geq 3$ is applied for all plates of the survey in order to identify spectral discontinuities. The precision of the technique has been determined by applying the algorithm to a grens plate containing the cluster of galaxies Abell 2670 previously studied by Sharples, Ellis, and Gray (1988). The precision obtained from the grens spectra, after removal of the cluster intrinsic velocity dispersion, gives a standard deviation of 2700 km sec$^{-1}$ for the cluster mean redshift.

We estimate the redshifts directly from the resulting list of breaks. An example of a grens galaxy spectrum can be found in BB. Redshifts are calculated from the shifted position of the 4000 Å stellar break. In most cases, it is the first break detected redward of 4000 Å and practically always the strongest in the spectrum.

Sometimes, the galaxies are bright enough for the detection of the shifted G-band ($\lambda_o \sim 4300$ Å). This serves to strengthen the redshift identification. When only one break is detected, it is always assumed that it is the 4000 Å break. Sources of confusion could arise from the Fe I blend ($\lambda_o \sim 3650$ Å) shifted redward of 4000 Å ($z > 0.1$) when this break is stronger than the 4000 Å stellar break. However, cases in which the Fe I blend are stronger than the 4000 Å break must be rare in early-type galaxies as shown by inspection of published galaxy distributions (for instance Pence 1976; Coleman et al. 1980).

Since the work of BB, we have implemented some improvements in the precision of the wavelength scale. The wavelength at position $i$ on each spectrum is determined from

$$\lambda_i = \lambda_o + \delta \Delta_i, \tag{1}$$

where $\lambda_o$ is the wavelength point of reference, $\delta$ is the wavelength dispersion and $\Delta_i$ is the distance of the position $i$ from the reference point ($\lambda_o$). The wavelength reference point used in the automated CFHT-Laval survey is the emulsion cutoff (green head) having a nominal value of $\lambda_o = 5356$ Å. The green head position is located for each spectrum with the break-finding algorithm (see Borra et al. 1987 for details).

The grens produces a dispersion which is not constant over the plate but varies along the dispersion direction. Furthermore, the atmospheric refraction adds an extra contribution to the nominal dispersion $\delta_o$ of the grens ($\delta_o = 945$ Å mm$^{-1}$). Both effects can be approximated by

$$\delta(\beta, \delta_r) = \beta(X - X_O) + \delta_O + \delta_r, \qquad (2)$$

where $\beta$ represents the position dependance of the dispersion with respect to the plate center ($X_O$) and $\delta_r$ the extra contribution of the atmospheric refraction to the nominal dispersion of the grens. Consequently, using a constant dispersion in (1) distorts the wavelength scale. This effect must be taken into account. From measures undertaken on the stellar Gunn and Stryker library of spectra (Gunn & Stryker 1983) with the break-finding algorithm, we have determined that the position of the 4000Å break is 3970±15 Å for stars with typical colors of galaxies. With both the nominal values of the green head and the 4000 Å break, we can derive $\delta(\beta, \delta_r)$ using bright stars over the plates to map equation (2). The dispersion $\delta(\beta, \delta_r)$ is given as the ratio of the 4000 Å - green head yardstick distance (1386 Å) to the distance measured on the spectrum, in mm, as determined from the break-finding algorithm. A simple plot of $\delta(\beta, \delta_r)$ vs ($X - X_O$) gives the constants $\beta$ and ($\delta_O + \delta_r$) to be applied to each plate in order to redefine the wavelength scale [equations (1) and (2)]. Using these constants, new positions are recalculated for every breaks. The one standard deviation error on $\delta(\beta, \delta_r)$ is ±3 Å. All redshifts are calculated using 3970 Å as the rest wavelength of the 4000 Å stellar break.

## 4 THE REDSHIFT DISTRIBUTIONS OF GALAXIES

Fig. 1 presents the redshift histograms for each individual field and for the co-added distributions in each of the 3 regions of sky (SGP, F and H). Fig. 2 shows the redshift distribution observed for NGP2. In order to eliminate binning effects, we have also cross-correlated the distributions with a boxcar of width $\Delta z = 0.01$: we sort the redshift distribution in ascending order, then count the number of galaxies inside a fixed window of $\Delta z =0.01$ from $z =0$ to $z =0.3$ in steps of 0.001 in redshift. This is equivalent to a detection technique to find structure, superposed on a smooth $z$ distribution, on a scale of $\Delta z =0.01$ (the window of $\Delta z =0.01$ is equal to the redshift precision). The results of the cross-correlations are shown as dotted lines in Fig. 1 and Fig. 2. The smooth curves represent the predicted distributions for a homogeneous universe ($\Omega_o=0.04$ and $H_o=100$ km s$^{-1}$ Mpc$^{-1}$). The models have been computed as described in Shank et al. (1984) with the difference that we use the Schechter parameters derived from the CFA survey ($\phi^*=0.020$ galaxies mag$^{-1}$, $M_{B(o)}^*=-19.2$ and $\alpha =-1.1$; de Lapparent, Geller & Huchra 1989). The fact that different morphological types have a break detectable to different limiting magnitudes has been included in the models. The assumption of a unifrom galaxy distribution is obviously a gross simplification that is not in agreement with the known clumpy galaxy distribution. We only add it to our figures to have a reference distribution. Because there are uncertainties (e.g. from the limitting magnitudes and incompletness), the total counts of the model are normalized to the *total number of observed* galaxies for each plate.

The most striking feature in the distributions shown in Fig. 1 and Fig. 2 is an apparent excess in the galaxy counts at $z \sim 0.1$ in 8 of the 10 individual fields. Interestingly, four of our plates are situated near the south and the north galactic poles, regions already studied by Broadhurst et al. (1990; hereafter referred to as BEKS) who also found peaks at $z \sim 0.1$. In the scope of these results, we discuss in the next section the statistical significance of the peaks seen in Fig.1 and 2.

## 5 DISCUSSION

### 5.1 Are the peaks spurious ?

The only systematic effect that we have identified and which could cause spurious peaks is contamination by stars misclassified as galaxies. Late-type stars usually have prominent G-band absorptions features (~ 4300 Å) which, if the stars are misclassified, result in spurious redshifts near $z \sim 0.1$ (see Fig. 2 in BB). Far from creating spurious peaks, all other sources of errors that we considered tend to washout existing peaks. The assessment of stellar contamination in the galaxy catalogue is thus critical. However, such contamination must be small in our case since we have checked visually the extendedness of all galaxies on plate and, for galaxies with borderline status, we have intersected the grens catalogue set with the COSMOS data. We give below three estimates of the contamination rate.

To estimate the contamination, we can make the reasonable assumption that the rate of stars misclassified as galaxies by our on-plate eyeball inspection is the same as for the automated methods, i.e. that eyeball evaluation is as efficient as automated methods. The rate of stars misclassified as galaxies is about 5% for both COSMOS and grens automated methods (Section 2 and Heydon-Dumbleton et al. 1989). Consequently, by intersecting two sets of galaxies (grens vs COSMOS or grens vs on-plate eyeball evaluation), the expected stellar contamination rate is reduced to 0.25%. As there typically are no more than 200 objects per plate with detected breaks at $z = 0.1$ in an interval of $\Delta z = 0.01$, we estimate the contamination rate per plate to be 0.5 star misclassified as a galaxy with $z \sim 0.1$.

We can estimate in a different independent way the stellar contamination rate. As stated in Section 2, there typically are 5 objects per plate classified as galaxies by the grens software that the eye inspection does not classify as certainly extended; 4 of these objects are typically confirmed as being galaxies in the COSMOS catalogue. This mean that ~20% of the borderline objects are likely to be stars.

Obviously, the list of objects classified as extended from on-plate inspection itself suffers, because of the subjectivity of the process, from errors as some objects classified as definitively extended by a first inspection could be classified as borderline by a second inspection. If we can estimate this error, we obtain a second estimate of the contamination rate. Based on many repeated eyeball inspections of the same list of galaxies over different periods of time, we find that, typically, there are 2 objects per plate that see their status changed from certain to borderline. Consequently, by applying the rate of 20% borderline objects being stars on these 2 objects, we obtain an expected number of 0.4 misclassified star per plate. Furthermore, our databases show that only ~40% of objects have detected breaks at $z=0.1$ in an interval of $\Delta z=0.01$, leading us to a final estimate of misclassified star per plate in each peak of 0.16; in good agreement with the previous estimate of 0.5 per plate.

Finally, to make sure that stellar contamination is kept to a minimum, we have made two other checks for all objects contained in the peaks, in addition to the previous criteria (visual checks on plates, overlap with COSMOS data for borderline objects). We have inspected by eye all digitized energy distributions and then checked the colors of these objects in a pseudo *U-B* vs *B-V* plot. Most galaxies have spectra significantly different from those of stars and can thus be distinguished by eyeball inspection; only a few faint objects cannot be classified from their noisy energy distributions. Also, most galaxies are characterized by an ultraviolet excess compared to the stellar population and are well separated in a color-color diagram (see Fig. 13 of Borra et al. 1987 for an example). Applying these two additional criteria, we found that, in the entire data set, all but 2 of the grens galaxies in the peaks are also confirmed as galaxy by at least two of the following characteristics: they are COSMOS galaxies, they have been identified as extended on the plates, they possess an ultraviolet excess compared to the stellar main sequence, their digitized energy distributions are typical of galaxies. Of the 2 objects which made exception, one is definitely extended on the grens plate, while the other is classed as a galaxy by COSMOS. This means that we should have < 2 contaminants in the survey and therefore < 0.3 per plate, in excellent agreement with the previous two estimates.

The previous three evaluations of the contamination rate agree quite well, indicating that the expected rate of misclassified stars in the peak is below ~ 0.5 per plate. That a contamination of 1 star per plate is unlikely is shown by another simple estimate. Consider that, to have 1 contaminating star per plate in a peak, the rate of stellar contamination has to be 0.5%. As the rate of stars misclassified as galaxies is 5% for the grens classifier (Section 2), the product of the misclassification rates of the two other criteria (see discussion in the previous paragraph) must then account for 10% contamination, therefore of order 30% for each individual criterion. It is certain that the misclassification rates of each of the criteria are well below the 30% rate. In summary, the only systematic effect that could cause spurious peaks, the contamination by stars misclassified as galaxies, is expected to be below ~ 0.5 per plate. Therefore, in the following section, we will neglect this small contribution to the count statistics.

We have also evaluated the accidental number of double stars expected per plate close enough to mimic an extended object (separation $\leq 1"$). We find that about 0.3 double stars per plate will have separation $\leq 1"$ (see Borra & Beauchemin 1992 for a discussion about blending statistics), 40% of which will have a G-band break, giving 0.12 contaminations in the peak, per plate.

The precision of the reference point for the wavelength calibration also is of concern. The wavelength cutoff position of the IIIa-J emulsion change with the colors of the objects. However, simulations show that the effect can be neglected for galaxies redshifted between $z=0.05$ and $z=0.2$. For these simulations, we have first redshifted an E0 galaxy template. Then we multiplied the template with the spectral response of a grens plate. Finally, we applied the break-finding algorithm to locate the green head. We found that the position of the green head varies by 14Å between $0.05 < z < 0.2$. Such a systematic effect cannot create the observed excess at $z \sim 0.1$ (or elsewhere). It is also important to stress that random errors will wash out existing peaks but will not create spurious ones.

Finally, there are 2 pieces of evidence that give us confidence in the reliability of the redshift catalogue to map large-scale structures:

(i) Fig. 2 shows the redshift distribution observed for NGP2. This plate is situated 18 arcminutes away from a region observed by Koo (1988) and therefore allows us a mean to verify the reliability of our results. Inspection of Fig. 2 reveals that galaxies in NGP2 are not smoothly distributed in redshift but instead present a sharp density excess at $z=0.123$ (a $\sim 4\,\sigma$ fluctuation). The distribution obtained by Koo also shows large peaks in the galaxy counts at certain redshifts separated by low density regions. In particular, his sharpest peak appears at $z=0.125$ with a FWHM of $\Delta z \sim 0.01$ (see also Fig. 1 of BEKS). This is exactly what emerges from Fig. 2. Furthermore, Koo stresses that half of all the galaxies in the sample have $B < 20$ and are concentrated in the $z=0.125$ peak.

(ii) In a recent paper, Guzzo et al. (1992) have presented a redshift survey for $\sim 100$ rich clusters centered at the SGP. They found a close correspondence between the maxima in their redshift catalogue and the periodic peaks detected in the BEKS survey. In particular, they found a peak at $z \sim 0.11$ located between $23^h.4 < \alpha < 1^h.2$ and $-40^o < \delta < -20^o$ and defining an elongated structure of $\sim 60 \times 140$ Mpc. The grens plates in the SGP ($\alpha=1^h, \delta=-19^o$) are situated on the edge of the Guzzo et al. $z \sim 0.1$ structure. The grens redshift distributions also show a peak at $z \sim 0.1$ in the SGP (see Fig. 1).

The fact that we have detected structures at $z \sim 0.123$ in NGP2 and at $z \sim 0.1$ in SGP, in agreement with the results of Koo et al. and Guzzo et al., gives support to the reliability of grens redshifts to map, with enough precision, the large-scale distribution of galaxies; at least for the NGP2 and SGP field.

### 5.2 Statistical significance of the peaks

We shall now try to obtain a measure of the statistical significance of the peaks, realizing at the outset that we suffer from small counts and that, consequently, cannot expect highly significant detection. Kaiser and Peacock (1991) have issued warnings against the overinterpretetion of pencil beams data. Although their warnings specifically addresses the pitfalls of power analysis and the periodicity found by BEKS, it is concerned with the effects of known clustering and is thus relevant to our discussion. We shall attempt to correct the statistical significance of our peaks for the known clustering of galaxies. Tables 2 and 3 give the count statistics for all peaks observed in Fig. 1 and 2 at $z \sim 0.1$. Counts have been estimated from the cross-correlated distributions since we are interested in the detection of peaks with $\Delta z \sim 0.01$. Columns in Tables 2 and 3 give respectively: the plate name, the redshift position of the peak, the observed/expected number of counts in the peak evaluated from Fig. 1 and 2, and the count excesses expressed in sigma units above the $n^{1/2}$ expectations for a random Poisson distribution. The expected number of counts is estimated from the adjacent bins. Tables 2 and 3 indicate that most of the peaks have $\sim 3\sigma$ signals above the background counts; a $2.5\sigma$ signal, when the expected value is 4 (the background is typically 4 counts at $z \sim 0.1$ in the survey), leads to a probability of $p \sim 0.02$ of being accidental. In most of the regions we probe (F, H and SGP), excesses are seen on at least 2 out of 3 plates. According to binomial statistics, the probability of obtaining at a same position 2 peaks with $p \sim 0.02$ on 3 plates is $\sim 0.02$, whatever the position in 17 bins. Table 3 also indicates that the co-added distributions inside each region (F, H and SGP) still exhibit peaks at $z \sim 0.1$ with $\sim 3\sigma$ signals. That a structure is present at $z \sim 0.1$ in each of the regions seems likely if one consider these statistics. However, in principle, Poisson statistics cannot be blindly applied to detect large-scale structure because galaxies are known to be clustered on a scale of a few Mpc; there will be an excess of variance in the cell counts compared to a random distribution. In presence of clustering described by the two-point correlation function, the count variance $\sigma^2$ in cells of volume V is given by

$$\sigma^2 = \langle (N - \langle N \rangle)^2 \rangle = \langle N \rangle (1+q) \qquad (3)$$

$$q = n\, V \langle \xi \rangle$$

$$\langle \xi \rangle = V^{-2} \int d^3r_1\, d^3r_2\, \xi(r_{12}), \qquad (4)$$

where $\langle N \rangle$ is the average number of galaxies ($n = \langle N \rangle V^{-1}$) and $q$ is the clustering correction to be applied to the random case (see, e. g., Maurogordato, Shaeffer & da Costa 1992). The variance reduces to the Poisson case when $\langle \xi \rangle$, the mean correlation term, is null. The correction factor $q$ can also be seen as the number of galaxies in a cell, in excess of random. Following BEKS, we approximate our pencil beams with a narrow cylinder of radius $R$ and total length

*D*. For this model, BEKS have derived the expression for the mean small-scale correlation when $\xi$ is described by the form $\xi = (r/r_o)^{1.8}$,

$$<\xi> = 2.24 \, r_o^{1.8} \, L^{-0.75} \, R^{-1.05}, \tag{5}$$

where we use $R = 2$ Mpc, $D = 675$ Mpc and $L = 30$ Mpc for the length of individual cell ($L \equiv$ redshift bins size). Which value of $r_o$ has to be used for our survey is not clear as early-type galaxies should dominate to some extend our sample. Using $r_o = 5$ Mpc (Shank et al. 1983), we obtain $q = 3.4$ and the significances of the peaks in Tables 2 and 3 should drop by about half their values. If we use $r_o = 10$ Mpc, value appropriate for a sample exclusively composed of elliptical galaxies, we obtain $q = 11.8$ and therefore the signals will be reduced by a factor ~ 3.5. Needless to say that the latter correction implies that all peaks seen in Fig. 2 and 3 are no longer significant. However, this evaluation applies only to data with error-free redshifts, which is not our case. In fact, our estimated redshift precision for individual galaxy, $\Delta z = 0.01$ ($\sigma_z \sim 30$ Mpc), is 3 to 6 times greater than the known clustering length ($5 < r_o < 10$ Mpc). This is a key point: such an uncertainty heavily smoothes any clumping due to the small-scale clustering and therefore significantly reduces the value of $\sigma^2$, the variance of the counts. We have performed simple Monte Carlo simulations to roughly estimate the effect of error measurements on the value of $\sigma^2$. We have randomly distributed 40 galaxies between $0.05 < z < 0.15$ (in this redshift range the observed counts and the selection functions are rather flat) with the particularity that, instead of distributing galaxy one by one, we place a clump of 4 galaxies at a time at the same position to mimic the clustering effect. Then, we compute the standard deviation in the cell counts ($\Delta z = 0.01$), with respect to the mean counts, before and after convolving the unbinned $z$ distributions with a Gaussian having $\sigma_z = 0.01$. We find that the standard deviation $\sigma$ of the counts drops by a factor of ~ 1.5 when smoothed by our redshift errors. Although this is a simple model, the exercise shows that our measurement errors must have an important effect on the expected variance. Exact simulations including redshift errors are more difficult to model, therefore, we have preferred to use directly our data to estimate the variance introduced by the small-scale galaxy clustering. To that aim, we have calculated for each plate the ratio of the variance to the mean count ($\sigma^2/<n>$) from the binned data between $0.05< z <0.15$ ($\Delta z =0.01$ bins). The value of $\sigma^2/<n>$ is expected to be equal to unity if counts are randomly distributed. Table 4 gives the $\sigma^2/<n>$ values obtained for each individual plate. The overall value for all plates combined, $\Sigma \sigma_i^2 / \Sigma <n_i>$ [which is also equal to the mean value $<(\sigma^2/<n>)>$], is 1.9 leading to a value of $q =0.9$. Therefore, the ~ $3\sigma$ signals in Table 2 and 3 will fall to $3\sigma/(1+q)^{1/2}=3\sigma/\sqrt{1.9} \sim 2\sigma$ [see equation (3)]. A $2\sigma$ signal represents an accidental probability of 0.05 when a random distribution has an expected value of 4 counts. The last column of Tables 2 and 3 give the significances of the peaks in standard deviation units after the $1/\sqrt{1.9}$ correction.

Let us now revise the probabilities of observing peaks like those in Fig. 1 and 2 with the new correction factor of $1/\sqrt{1.9}$. Signals at $z \sim 0.1$ in the F fields are still present in all 3 plates with ~ $2\sigma$ significances. The probability of obtaining by chance at a same position 3 peaks with $p \sim 0.05$ on 3 plates in 17 bins is 0.002, a low probability value. The $4.5/\sqrt{1.9}=3.3\sigma$ signal at $z \sim 0.1$ in SGP1 is at the borderline of significance. Finally, in the H fields, the accidental probability of obtaining at a same position 2 peaks with $p \sim 0.05$ ($2\sigma$) on 3 plates in 17 bins is marginal at best with 0.07. However, the co-added distributions inside the H region, as well as F and SGP regions (see Table 3), still exhibit $> 2\sigma$ signals at $z \sim 0.1$ after the $1/\sqrt{1.9}$ correction with $2.5\sigma$, $2.5\sigma$ and $2.1\sigma$ signals respectively. The $4\sigma$ Poisson signal at $z =0.123$ in NGP2 reduces to $2.9\sigma$ after clustering correction. Finally, the probabilities of obtaining in 17 bins 3 individual $2\sigma$ signals ($p \sim 0.02$) at a same position over 3 different regions is only $\sim 10^{-4}$ (we consider F, H and SGP having their peaks at the same position $0.102 \leq z_{peak} \leq 0.111$).

The correction that was applied render this statistical discussion somewhat unsatisfactory, but is the best we can do under the circumstances. After estimating directly from our data the correction for small-scale galaxy clustering, it seems that a $\sim 2\sigma$ signal still stands out at $z \sim 0.1$ in each of the 4 regions we probe. Signals $\sim 2\sigma$ are marginal, especially considering the uncertainties in the correction applied. The significance is however enhanced because these peaks are all detected at $z \sim 0.1$. The confidence in our detection is greatly enhanced by the detection by independent workers of peaks at $z = 0.1$ in fields coinciding or near our fields. BEKS find peaks in a field that coincides with our NGP field and a second one 9 degrees away from our SGP fields. Recently, Vettolani et al. (1993) have presented the preliminary results of an extensive deep ( to bJ = 19.4) redshift survey. Their histogram of 670 galaxy redshifts obtained in a 22 X 1.5 ° strip of sky centered at an hour angle of zero hours and a declination of - 40 °, a field only 15 degrees from our SGP fields, shows a strong peak at z =0.1. Furthermore, redshifts obtained as part of observation of the cluster of galaxies Abell 151 by Proust et al. (1992) show the cluster and a void followed by an isolated clump of galaxies at z=0.1. This field is only 2 degrees away from our SGP field.

## 6  CONCLUSION

The peaks observed at $z \sim 0.1$ may have two origins. First, they may be due to some selection effects, systematic errors, statistical fluctuations or combinations of the above. Second, our pencil beams sample large-scale superstructures.

The first explanation is always possible but contamination of the catalogue by misclassified stars is the only systematic effect that we could identify. We were very careful to minimize this and our estimate of the contamination indicates that it does not contribute significantly to the peaks. On the other hand, it is surprising that structures can stand out in presence of measurement errors as large as the bin size used to detect them, unless such structures are thin. But it seems that this can be the case since the largest known structure in the universe, the Great Wall, is indeed thin having only 5 Mpc thickness (Ramella, Geller & Huchra 1992).

Assuming that our pencil beams have detected large scale structures, it is legitimate to wonder how large they are and to compare them with our present knowledge of large scale structure in the universe. On the other hand, any speculation must be tempered by the sparse sampling of our pencils beams; spectroscopy between fields is needed to follow-up on what our data may suggest.

At $z \sim 0.1$, the separation of 9 degrees in SGP between the grens and the BEKS fields represents 45 Mpc ($H_o$=100 km s$^{-1}$ Mpc$^{-1}$), comparable to the dimensions of known superclusters. Sets of 3 plates inside each region subtend maximum angles of $\sim$ 3 deg or 15 Mpc, which is 15-30% the dimension of superclusters (Oort 1983). It is thus reasonable to suggest that the peak in the SGP grens fields and the one in BEKS belong to the same structure, also detected by Guzzo et al. (1992). We also find that the typical number of grens galaxies is of the same order as observed by BEKS in SGP when field surfaces are normalized and counts are extrapolated for the difference of $\sim$ 2 mag between the two surveys [we use $\gamma$=0.5 ($\gamma$= d log $N$ / d$m$ )]. If the peaks seen by Vettolani et al. belongs to the same structure, it would have dimensions of the order of 30 degrees, corresponding to 150 MPc similar to the Great Wall (Geller & Huchra 1989).

Do the peaks in the F fields belong to the same structure as the SGP and the BEKS fields so that our pencil-beams are probing a superstructure extending ~ 45 degrees across the sky ? First one must note that, unlike the NGP fields, there is no independent observational evidence of the reality of the peaks. If we are probing structures similar to the Great Wall (Geller & Huchra 1989), Ramella et al. (1991) show that for a beam-width of the dimension of a grens plate, i.e. ~ 4 Mpc at $z$ =0.1; there is ~ 30% of chance of detecting such a structure with an excess of 5 counts. In fact, we have detected an excess of the same order on 80% of our plates. A separation of 45 degrees across the sky, the distance between the F and SGP fields, would represent a structure of ~ 220 Mpc at $z$ =0.1, which is ~ 1.5 times larger than the Great Wall. The separation of 50 degrees between the H and NGP2 fields is of the same order as the separation between F and SGP, leading to the same speculation that they also sample a structure of order 250 Mpc. These hypotheses would not be grossly out of line with our present knowledge of large scale structures. Spectroscopy between fields is needed to follow-up on this.

.## AKNOWLEDGEMENTS

...

The authors thank H. T. MacGillivray who provides us COSMOS data with the stars/galaxies separation already performed. We also wish to thank the David Dunlap Observatory for the use of their PDS microphotometer. This research has been supported by the Natural Sciences and Engineering Research Council of Canada.

**FIGURE CAPTIONS**

Fig. 1. The redshift distributions for each plate and for the co-added distributions in each of the 3 regions of sky. The dotted lines show the results obtained from a cross-correlation of the redshift distribution with a boxcar of $\Delta z = 0.01$ (see Section 4). The smooth curve represents the predicted distribution for a homogeneous universe normalized to the total number of observed galaxies.

Fig. 2. The redshift distributions observed for NGP2. Symbols are the same as Fig. 1. There is a sharp peak at $z = 0.123$. This plate is situated 18' away from a region observed by Koo (1988) where an excess of galaxies at $z = 0.125$ have been found.

Table 1. Survey fields.

| Field name | R.A. (1950) | Dec. | Limiting $B$ magnitude |
|---|---|---|---|
| F1 | 03:59:47 | -13:58:38 | 21.6 |
| F2 | 03:59:52 | -14:58:49 | 21.7 |
| F3 | 03:59:50 | -15:58:48 | 21.8 |
| H3 | 09:26:48 | +18:00:17 | 21.8 |
| H5 | 09:26:43 | +16:00:33 | 21.7 |
| H6 | 09:26:43 | +15:00:17 | 21.8 |
| NGP2 | 13:04:21 | +29:43:29 | 21.2 |
| SGP1 | 01:00:00 | -18:00:00 | 21.0 |
| SGP2 | 01:00:00 | -19:00:00 | 21.7 |
| SGP3 | 01:00:00 | -20:00:00 | 21.7 |

26Table 2. Count statistics for all peaks observed in Fig. 1 and 2.

| Field | $z_{n=max}$ | Counts observed ($N_o$) / Counts Expected ($N_e$) | Excess $(N_o - N_e) / N_e^{1/2}$ | Excess $/\sqrt{1.9}$ [†] |
|---|---|---|---|---|
| F1   | 0.105 | 9 / 3  | 3.46 | 2.5 |
| F2   | 0.105 | 9 / 4  | 2.50 | 1.8 |
| F3   | 0.108 | 9 / 4  | 2.50 | 1.8 |
| H3   | -     | -      | -    | -   |
| H5   | 0.111 | 9 / 3  | 3.46 | 2.5 |
| H6   | 0.115 | 11 / 5 | 2.68 | 1.9 |
| NGP2 | 0.123 | 10 / 3 | 4.00 | 2.9 |
| SGP1 | 0.105 | 13 / 4 | 4.50 | 3.3 |
| SGP2 | 0.095 | 9 / 5  | 1.80 | 1.3 |
| SGP3 | -     | -      | -    | -   |

[†] Corrected for small-scale clustering, see Section 5.2

Table 3. Count statistics for all peaks observed per region.

| Field | $z_{n=max}$ | Counts observed ($N_o$) / Counts Expected ($N_e$) | Excess $(N_o-N_e) / N_e^{1/2}$ | Excess $/\sqrt{1.9}$ [†] |
|---|---|---|---|---|
| Σ F | 0.105 | 24 / 12 | 3.46 | 2.5 |
| Σ H | 0.111 | 21 / 10 | 3.48 | 2.5 |
| Σ SGP | 0.102 | 30 / 18 | 2.83 | 2.1 |
| NGP2 | 0.123 | 10 / 3 | 4.00 | 2.9 |

[†] Corrected for small-scale clustering, see Section 5.2

Table 4. Variance-to-mean ratios observed between 0.05< $z$ <0.15 in $\Delta z$ =0.01 bins.

| Field | $\sigma^2$ | $<n>$ | $\sigma^2/<n>$ |
|---|---|---|---|
| F1 | 7.57 | 2.3 | 3.29 |
| F2 | 6.04 | 3.6 | 1.68 |
| F3 | 3.88 | 3.9 | 1.00 |
| H3 | 1.60 | 2.4 | 0.67 |
| H5 | 3.73 | 2.2 | 1.70 |
| H6 | 14.1 | 5.5 | 2.56 |
| NGP2 | 6.32 | 2.9 | 2.18 |
| SGP1 | 10.5 | 4.1 | 2.57 |
| SGP2 | 7.38 | 4.6 | 1.67 |
| SGP3 | 10.3 | 6.4 | 1.6 |

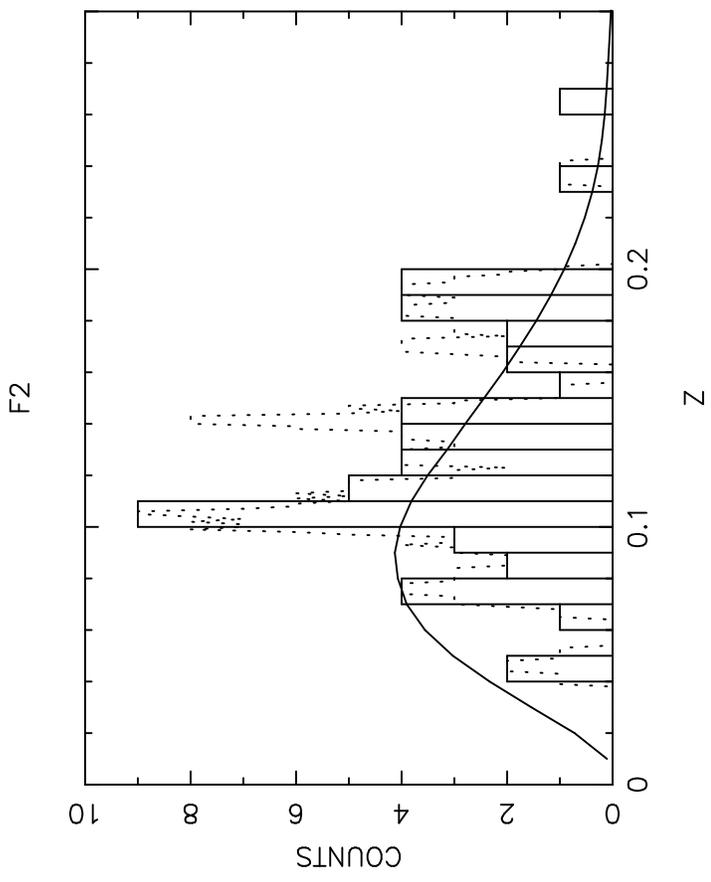
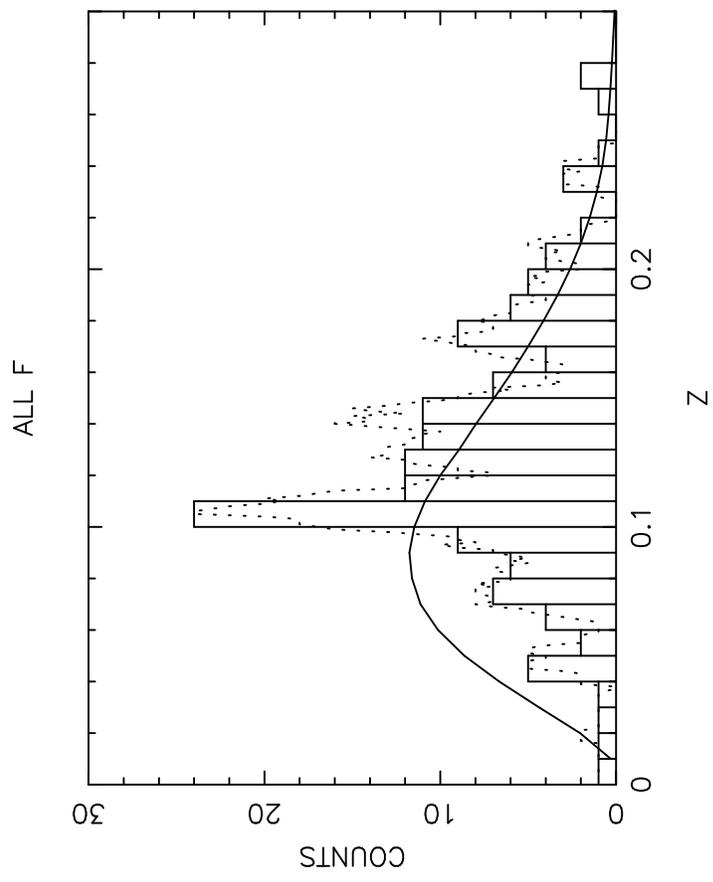
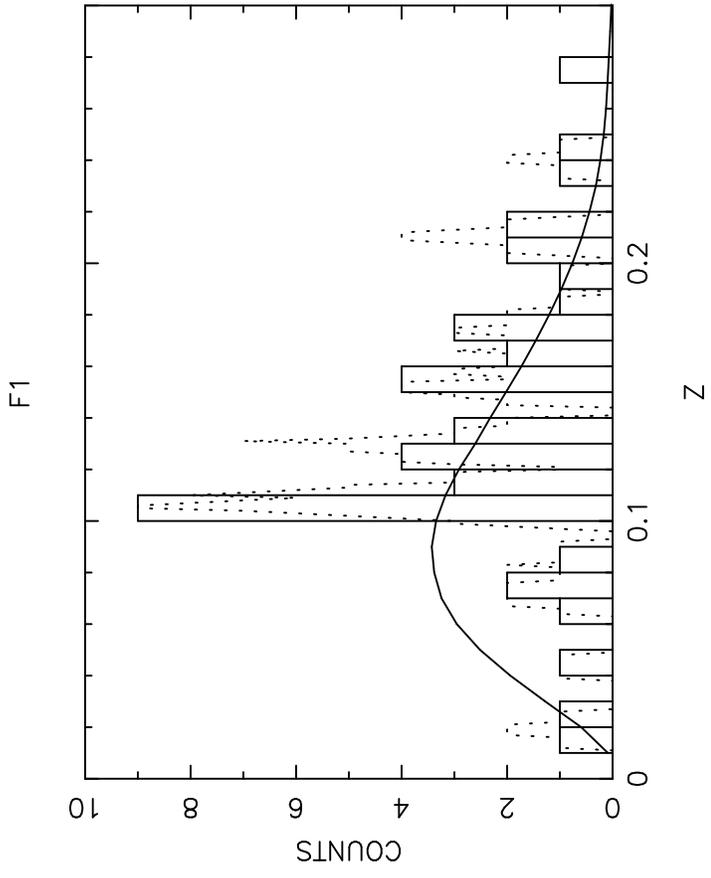
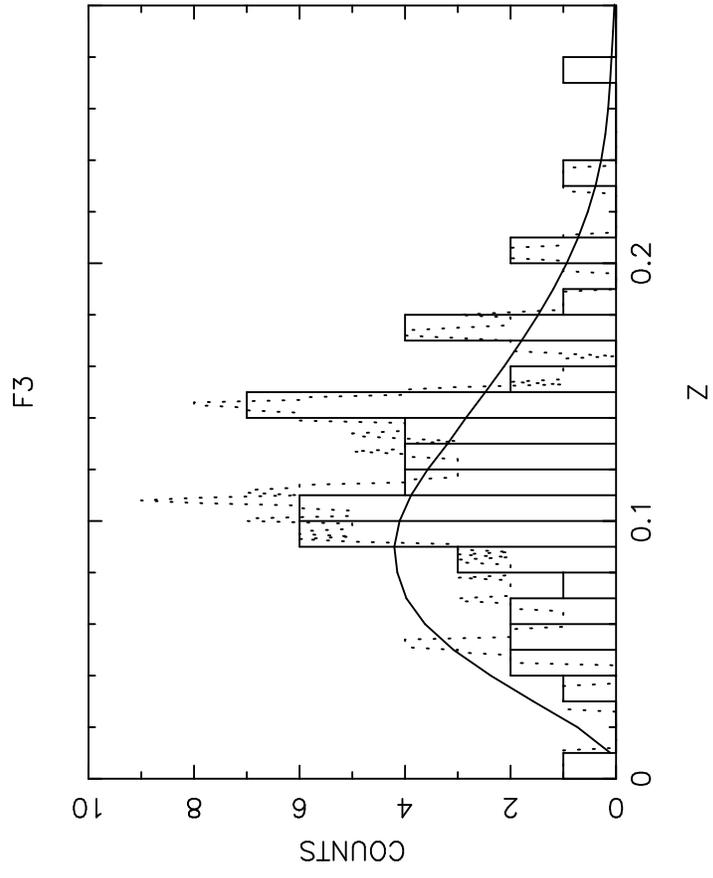

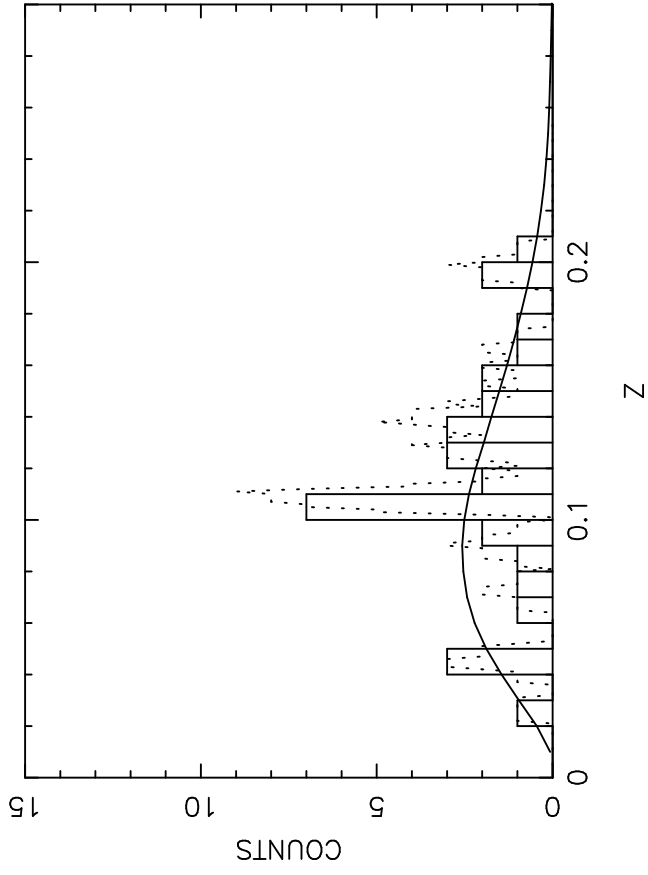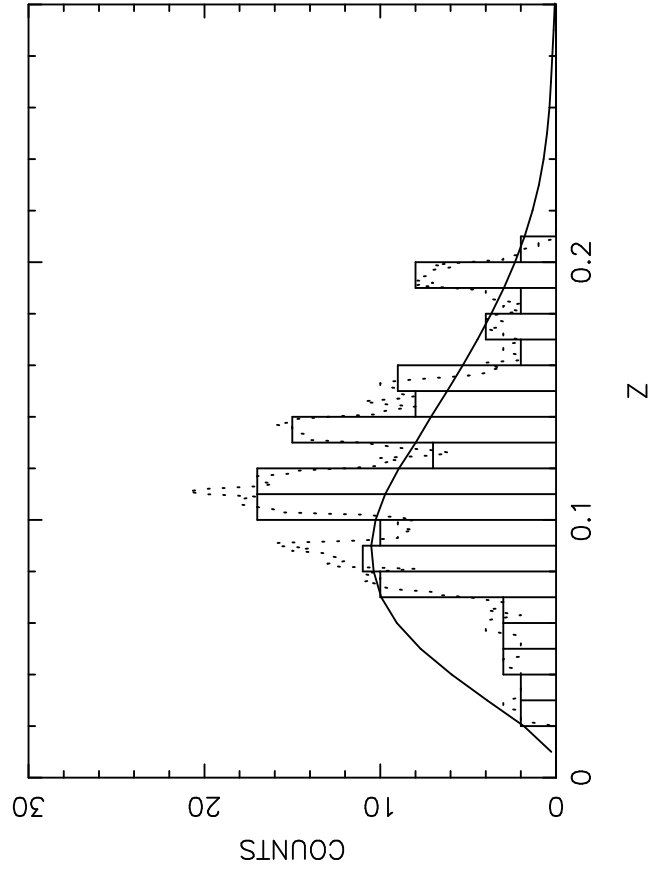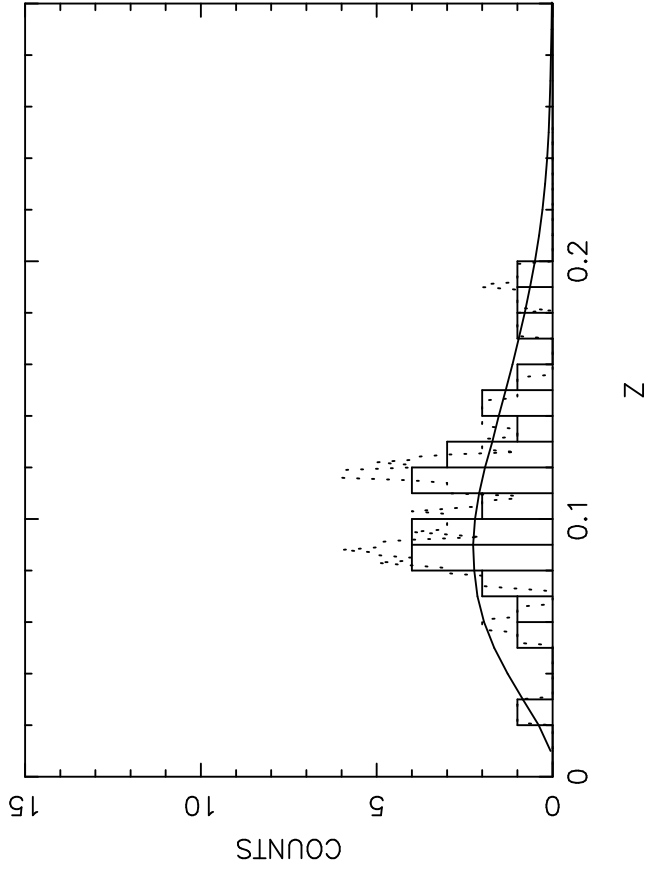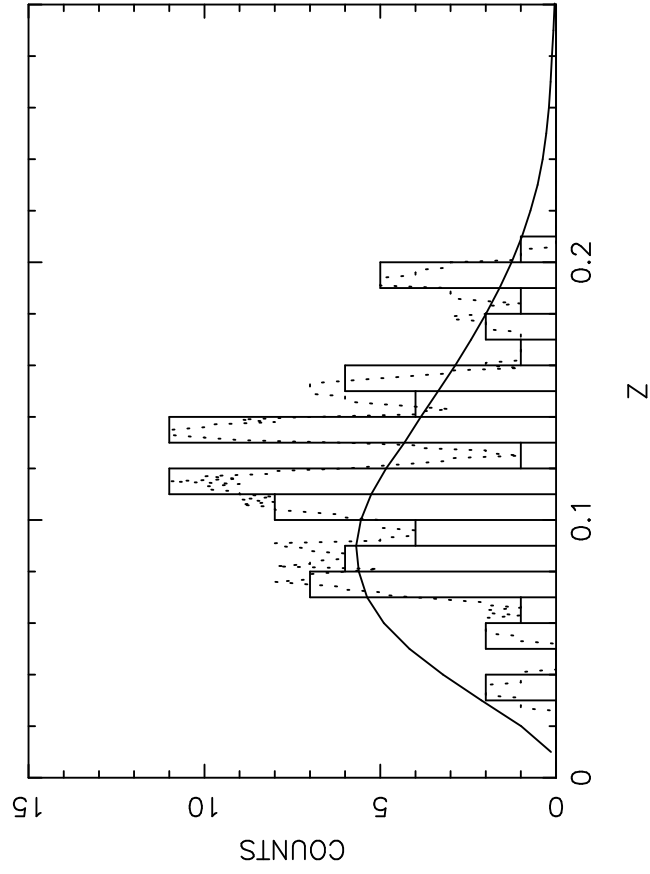

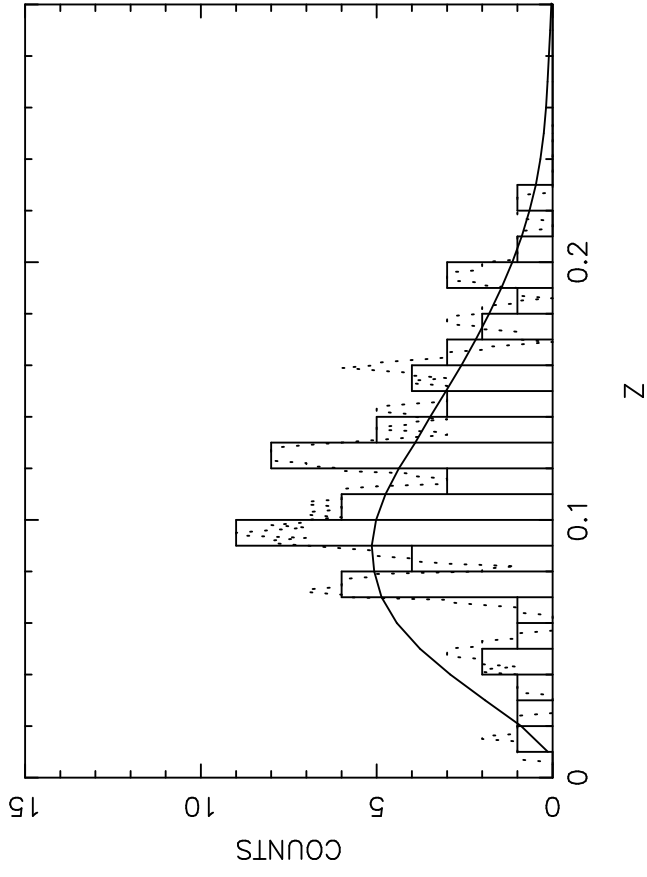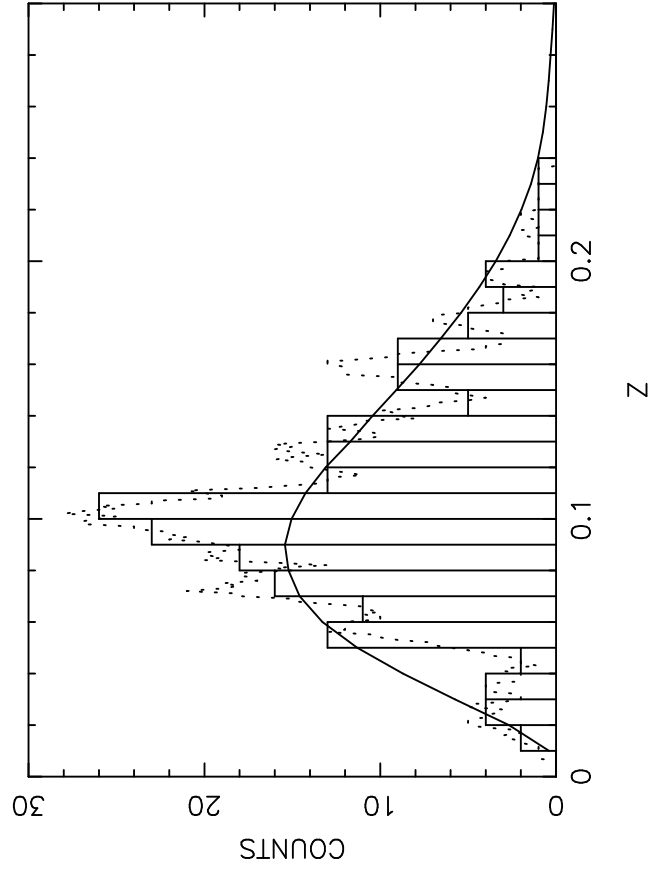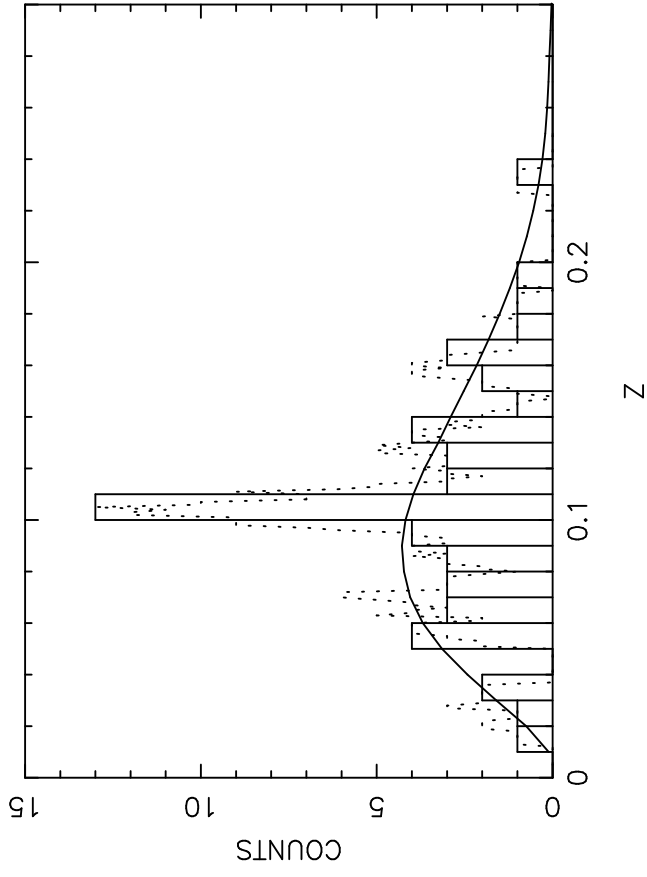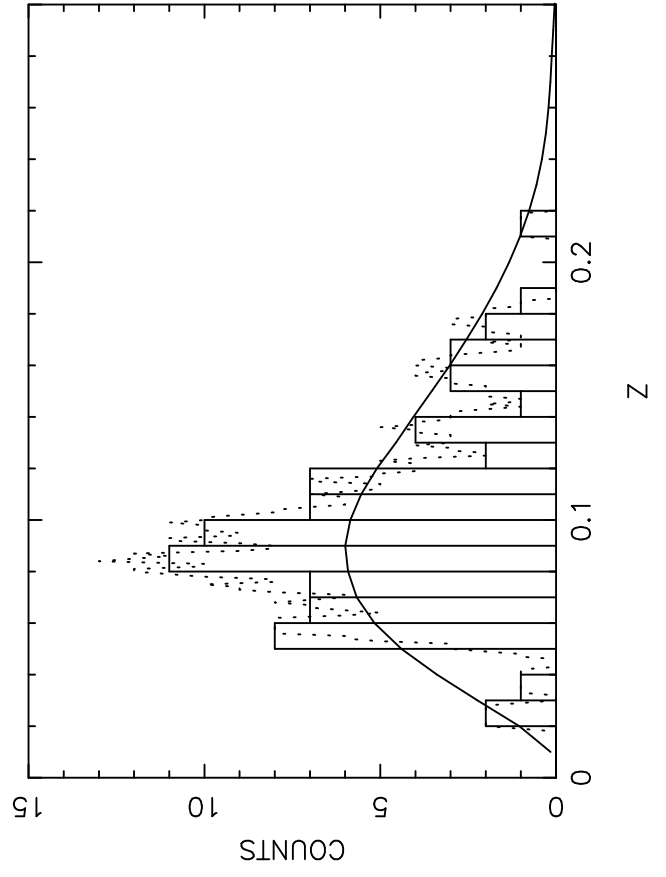

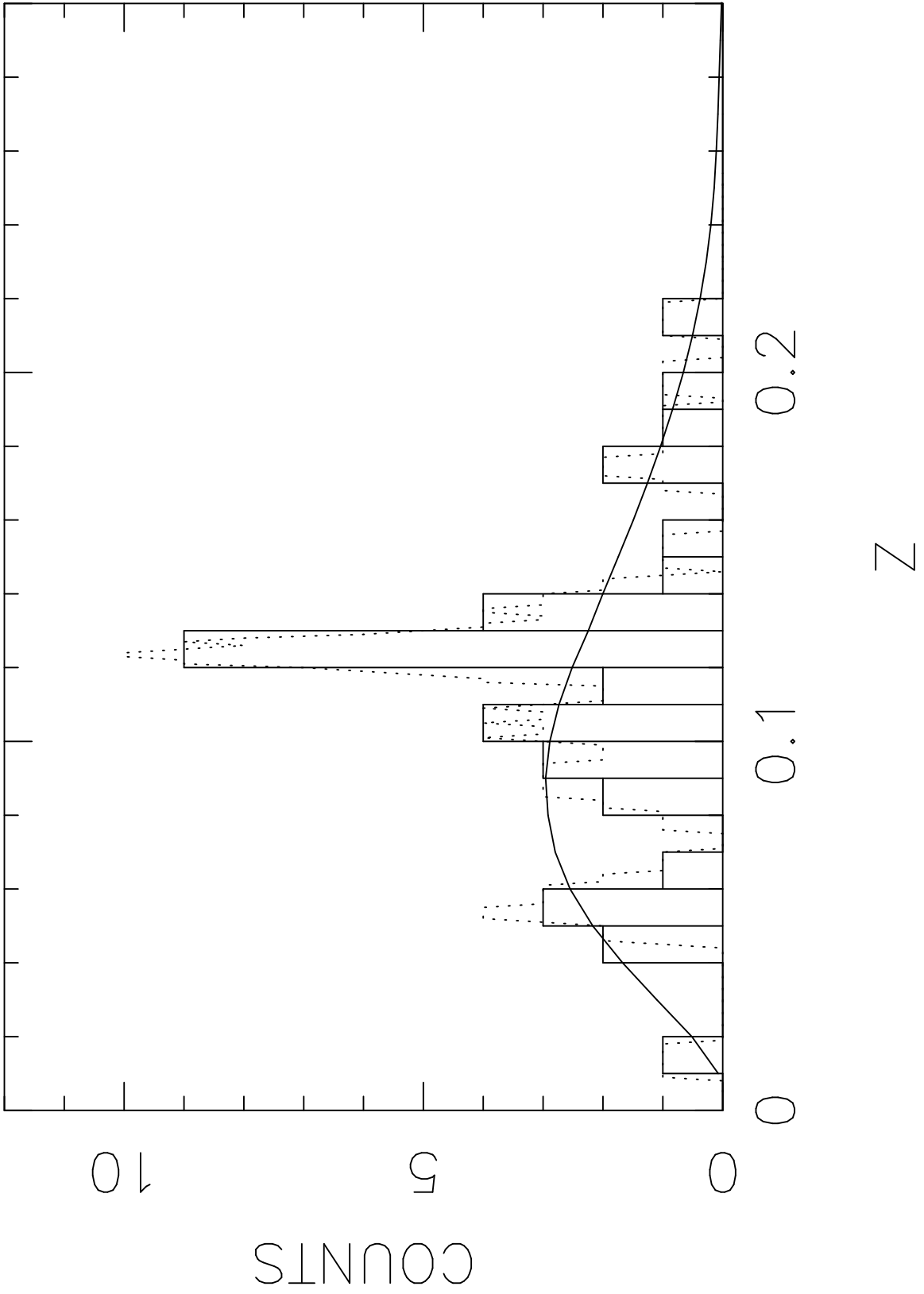